\documentclass[journal]{IEEEtran}

\usepackage[font=footnotesize]{caption}
\usepackage{subfig}

\ifCLASSINFOpdf
  \usepackage[pdftex]{graphicx}
  \DeclareGraphicsExtensions{.pdf}
\else
\fi

\usepackage{cite}
\usepackage{amsmath}
\usepackage{multirow}
\usepackage{array}
\usepackage{color}
\usepackage{url}
\usepackage{soul}
\usepackage{nicefrac}
\usepackage{xfrac}

\begin{document}

\title{LSTM-Based ECG Classification for Continuous Monitoring on Personal Wearable Devices}
%

\author{Saeed~Saadatnejad,~
        Mohammadhosein~Oveisi,~
        and~Matin~Hashemi
        
{\color{blue} 
\begin{flushleft}
\footnotesize 
This article is published. Please cite as: S. Saadatnejad, M. Oveisi, M. Hashemi, "LSTM-Based ECG Classification for Continuous Monitoring on Personal Wearable Devices", IEEE Journal of Biomedical and Health Informatics (JBHI), 2019. 
\end{flushleft} }
        
\thanks{Authors are with the Learning and Intelligent Systems Laboratory, Department of Electrical Engineering, Sharif University of Technology, Tehran, Iran. 
Webpage: http://lis.ee.sharif.edu, E-mail: saeedsa@ee.sharif.edu, oveisi@ee.sharif.edu, matin@sharif.edu (corresponding author).}
}

%
%

\markboth{IEEE Journal of Biomedical and Health Informatics}%
{IEEE Journal of Biomedical and Health Informatics}
%



\maketitle

\begin{abstract}
\emph{Objective}: A novel ECG classification algorithm is proposed for continuous cardiac monitoring on wearable devices with limited processing capacity. 
\emph{Methods}: The proposed solution employs a novel architecture consisting of wavelet transform and multiple LSTM recurrent neural networks (Fig. \ref{fig:alg}). 
\emph{Results}: 
Experimental evaluations show superior ECG classification performance compared to previous works. Measurements on different hardware platforms show the proposed algorithm meets timing requirements for continuous and real-time execution on wearable devices. 
\emph{Conclusion}: In contrast to many compute-intensive deep-learning based approaches, the proposed algorithm is lightweight, and therefore, brings continuous monitoring with accurate LSTM-based ECG classification to wearable devices. 
\emph{Significance}: The proposed algorithm is both accurate and lightweight. 
The source code is available online \cite{SourceCode}. 
\end{abstract}


\section{Introduction}
\label{sec:intro}

\IEEEPARstart{C}{ardiovascular} diseases (CVDs) such as myocardial infarction, cardiomyopathy and myocarditis are the leading causes of death in the world. An estimated $17.7$ million people died from CVDs in $2015$, representing $31\%$ of all global deaths reported by the World Health Organization \cite{CVDwho}. Cardiac arrhythmias are among the most important CVDs. 

Electrocardiogram (ECG) signal represents electrical activities of the heart and is widely used in detection and classification of cardiac arrhythmias. A trained cardiologist can detect arrhythmias by visually inspecting the ECG waveform. 
However, arrhythmias occur intermittently, especially in early stages of the problem. Hence, it is difficult to detect them in a short time window of the ECG waveform. 
Therefore, continuous monitoring of patients' heartbeats in daily life is crucial to arrhythmia detection  \cite{jbhi_dsp_qrs_2018}. 

Wearable devices provide a platform for this purpose \cite{jbhi_dsp_qrs_2018}. 
Our approach is to locally execute the ECG classification algorithm on patients' personal wearable devices. 
Local execution allows for continuous operation regardless of the network speed and availability. In addition, it allows data to stay on the wearable device and hence avoids privacy issues of cloud-assisted processing. 
Our approach is different from offline processing of stored ECG signals, or remote processing on powerful cloud servers \cite{jbhi_cloud_2014, jbhi_cloud_2015}. 

Continuous monitoring on wearable devices require the automated ECG classification algorithm to be both \emph{accurate} and \emph{light-weight} at the same time. This forms our main focus in this work. Note that wearable devices have small and low-power processors which are much slower compared to desktop and server processors.

Many previous algorithms are based on morphological features and classical signal processing techniques \cite{jbhi_dsp_cluster_2018, Chazal2004, Minami1999, Hermite2000, Shyu2004, Inan2006, Melgani2008, Coast1990, Hu1997, Chazal2006, Jiang2007, Ince2009}. 
Since the ECG waveform and its morphological characteristics, such as the shapes of QRS complex and P waves, significantly vary under different circumstances and for different patients, the fixed features employed in such algorithms are not sufficient for accurately distinguishing among different types of arrhythmia for all patients \cite{Hoekema_2001, Kiranyaz2016}.

To extract the features automatically and increase the heartbeat classification accuracy, deep-learning based algorithms including deep convolutional neural networks and recurrent neural networks have recently been proposed \cite{Kiranyaz2016, Ng2017, Kachuee2018, 2018_dcnn_1, GR_NN_2017}.

\begin{figure}[tp]
	\centering
	\includegraphics[width=1.0\columnwidth]{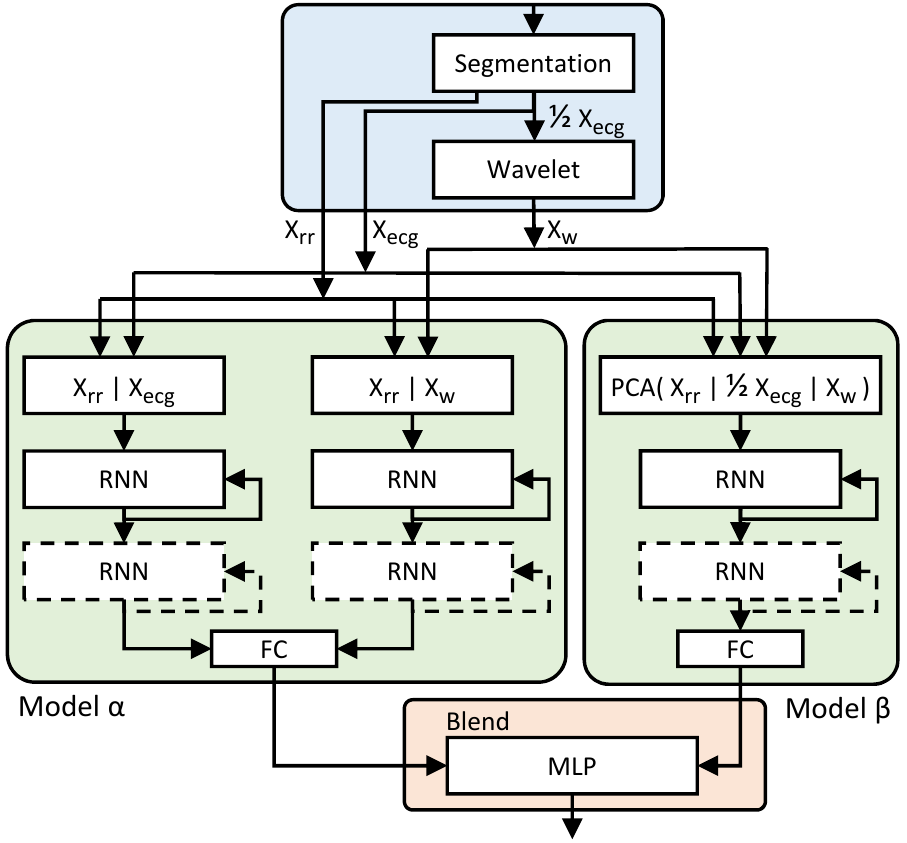}
	\vskip -1mm
	\caption{Overall view of the proposed algorithm.}
	\vskip -2mm
	\label{fig:alg}
\end{figure}

\begin{figure}[tp]
	\centering
	\includegraphics[width=1.0\columnwidth]{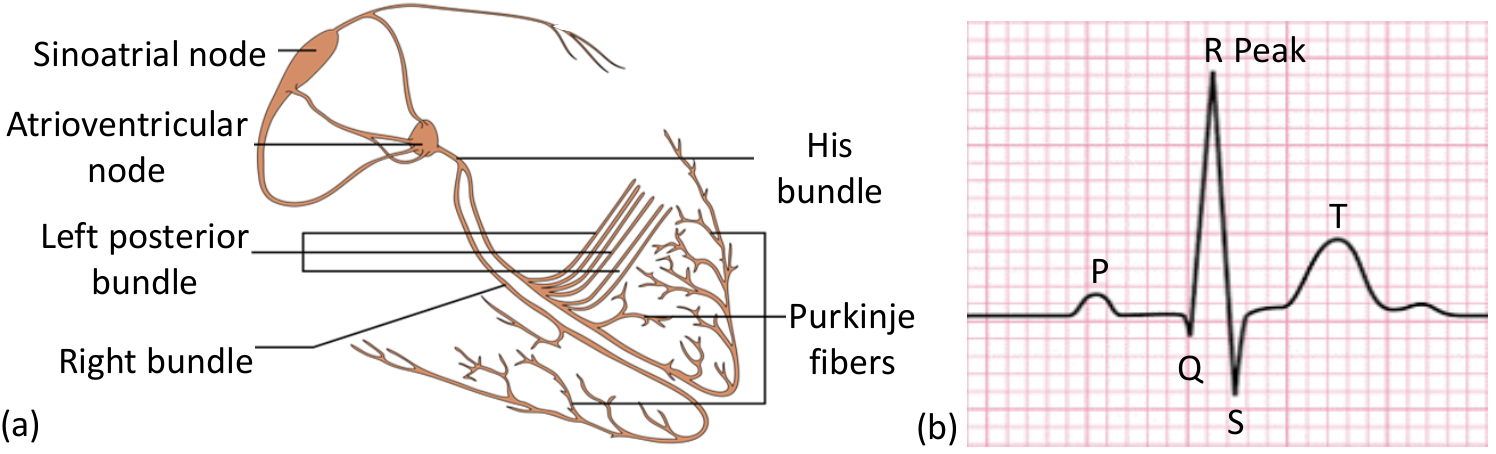}
	\vskip -1mm
	\caption{(a) Electrical system of the heart. (b) ECG waveform \cite{Heart,Heart2}.}
	\vskip -2mm
	\label{fig:heart}
\end{figure}

This paper proposes a novel ECG classification algorithm based on LSTM recurrent neural networks (RNNs). An overall view of the algorithm is shown in Fig.~\ref{fig:alg}. 
The proposed algorithm employs RNNs because the ECG waveform is naturally fit to be processed by this type of neural network. The reason lies within the electrical conduction system of the heart which is shown in Fig.~\ref{fig:heart}. 
The sinoatrial node generates a pacemaker signal which travels through internodal pathways to the atrioventricular node. The conduction slows through the atrioventricular node which causes a time delay. Next, the signal travels through the bundle of His to the heart apex and the Purkinje fibers, then finally to the ventricles \cite{Heart}. 
The above sequence of electrical activities are reflected into the ECG waveform (Fig.~\ref{fig:heart}), and therefore, temporal dependencies naturally exist in this waveform. 
RNNs capture such temporal dependencies in sequential data more efficiently compared to other types of neural networks. 

As shown in Fig.~\ref{fig:alg}, the proposed algorithm employs both LSTM recurrent neural networks and classical features, i.e., wavelet, at the same time. The additional features help to better capture the patterns in the ECG waveform. 
In addition, the proposed algorithm merges the arrhythmia predictions from small LSTM models as opposed to constructing one large model. 
See models $\alpha$ and $\beta$ in Fig.~\ref{fig:alg}. 
The total computational costs for executing multiple smaller LSTM models is lower than one larger LSTM model.

As a result, in contrast to many previous deep-learning based approaches which are computationally intensive \cite{Ng2017,Kachuee2018,2018_dcnn_1,GR_NN_2017}, the proposed algorithm increases the classification accuracy without significantly increasing the computational costs. Hence, it brings continuous monitoring with accurate LSTM-based ECG classification to personal wearable devices.

Experimental results show effectiveness of the proposed solution. Reporting the accuracy of ECG classification algorithms has been standardized by the Association for the Advancement of Medical Instrumentation (AAMI) \cite{AAMI}. 
Our proposed algorithm is evaluated using the same ECG signals that were employed in the previous works that conform to this standard. 
Experimental evaluations demonstrate that the proposed algorithm has superior classification performance compared to such methods. 
For instance, $F_1$ score is $3.3\%$ and $15.5\%$ higher in classifying ventricular ectopic beats (VEB) from non-VEBs and supraventricular ectopic beats (SVEB) from non-SVEBs, respectively. Note that SVEB detection is considered to be more difficult than VEB detection.

Computational requirements of the proposed algorithm is evaluated as well. Empirical measurements on small and low-power hardware platforms show that the proposed algorithm meets timing requirements for continuous and real-time execution on such platforms. 

Previous works have lower classification performance \cite{Chazal2004, Minami1999, Hermite2000, Shyu2004, Inan2006, Melgani2008, Coast1990, Chazal2006, Hu1997, Jiang2007, Ince2009, Hoekema_2001, Kiranyaz2016}, are not suitable for continuous execution on wearable devices due to high computational intensity \cite{Ng2017,Kachuee2018,2018_dcnn_1,GR_NN_2017}, do not include all the standard AAMI classes \cite{Chauhan2015, alexnet2017, Zhou2017, cbm2018_wavelet_lstm, cbm2018_cnn_lstm, 2018_dcnn_v}, or focus on other problems related to processing of ECG signals \cite{biometric2015, Salloum2017, crnn2017, ETH_RNN2017, Ubeyli2009, Ubeyli2010, af2017cnn, af2018, jbhi_cnn_af_2018, cinc2017, Lipton2017}.

Detailed comparisons with all the related works are presented in Section \ref{sec:relworks}. The proposed algorithm and its training procedure are discussed in Sections \ref{sec:alg} and \ref{sec:train}. The experimental results and discussions are presented in Sections \ref{sec:exp} and \ref{sec:exp:dis}. Concluding remarks are presented in Section \ref{sec:conc}.

\section{Related Works}
\label{sec:relworks}

Many previous ECG classification algorithms are mainly focused on signal processing techniques including extraction of morphological features \cite{Chazal2004}, frequency domain analysis \cite{Minami1999}, Hermite function decomposition \cite{Hermite2000}, wavelet transform \cite{Shyu2004,Inan2006}, support vector machines \cite{Melgani2008} and hidden Markov models \cite{Coast1990}. 
Hu et. al \cite{Hu1997} proposed a mixture of experts method for patient-adaptable heartbeat classification. 
Chazal and Reilly \cite{Chazal2006} proposed a personalized heartbeat classification algorithm based on linear discriminant analysis on ECG morphology and timing interval features. 
Jiang and Kong \cite{Jiang2007} proposed a block-based neural network algorithm, and Ince et. al. \cite{Ince2009} proposed particle swarm optimization for artificial neural networks, both for patient-specific heartbeat classification. 
Compared to the above solutions, the proposed algorithm achieves higher classification performance.

Recent approaches have focused on deep learning. Kiranyaz et. al. \cite{Kiranyaz2016} proposed a one-dimensional convolutional neural network algorithm. Both the above and the proposed methods meet timing requirements, but our proposed method achieves higher classification performance, especially in SVEB detection. In \cite{Kiranyaz2016}, a heartbeat trio is fed into the network in order to capture the effect of nearby heartbeats in classifying the current heartbeat. This overhead is not necessary in our method since the LSTM cells capture temporal dependencies automatically and more efficiently. In addition, our proposed solution combines the arrhythmia predictions from small LSTM models as opposed to constructing one large model. 

Rajpurkar et. al. \cite{Ng2017} proposed a much deeper CNN. This computationally intensive algorithm is designed for a different problem namely classifying ECG signals into rhythms such as Sinus and Bigeminy. It consists of $34$ layers and is not suitable for execution on wearable devices due to its very long execution time. It is about $10,000$X slower than the proposed method. 
Kachuee et. al. \cite{Kachuee2018} proposed a deep CNN algorithm with $11$ layers which is less accurate and about $100$X slower compared to the proposed method. 
Jun et. al. \cite{2018_dcnn_1} proposed another CNN algorithm with $11$ layers but their convolutions are two-dimensional and hence much more computationally intensive than one-dimensional convolutions. 
In contrast, our proposed method is designed from ground up to be lightweight, and hence, meets timing requirements for continuous execution on wearable devices with limited processing capacity.

A general regression neural network was proposed in \cite{GR_NN_2017} for classification of long-term ECG signals. This algorithm is designed for offline processing and requires the entire recorded data. The algorithm is accelerated on high-performance GPUs in order to reduce the execution time. 
Teijeiro et. al. \cite{jbhi_dsp_cluster_2018} proposed an ECG clustering algorithm that requires the entire recorded data. 
In contrast, our proposed algorithm is able to classify real-time ECG signals.

There are many other deep-learning based ECG classification methods in the literature that do not comply with AAMI standards and hence are not directly comparable with the proposed solution. 
For instance, many only consider a selected subset of the standard classes \cite{Chauhan2015, alexnet2017, Zhou2017, cbm2018_wavelet_lstm, cbm2018_cnn_lstm, 2018_dcnn_v}, which makes the design and training of neural networks much simpler because not all the challenging cases are included. 
The proposed solution fully complies with AAMI standards \cite{AAMI}, the results are reported based on the standard and openly available MIT-BIH dataset \cite{MIT-BIH} and all standard classification metrics have been calculated and reported.

Deep learning has also been applied to other problems related to analysis of ECG signals, for instance, ECG-based biometrics \cite{biometric2015,Salloum2017}, detecting atrial fibrillation (AF)  \cite{crnn2017, ETH_RNN2017, Ubeyli2009, Ubeyli2010, af2017cnn, af2018, jbhi_cnn_af_2018} which is normally based on CinC Challenge 2017 dataset \cite{cinc2017}, and diagnosis based on hospital records \cite{Lipton2017}. 

\section{Proposed Algorithm}
\label{sec:alg}

Fig. \ref{fig:alg} presents an overall view of the proposed algorithm. First, the incoming digitized ECG samples are segmented into heartbeats and their RR interval features and wavelet features are extracted (Sections \ref{sec:alg:rpeak} and \ref{sec:alg:wavelet}). Next, the ECG signal along with the extracted features are fed into two RNN-based models which classify every heartbeat (Sections \ref{sec:alg:rnn} and \ref{sec:alg:cells}). The two outputs are then blended to form the final classification for every heartbeat (Section \ref{sec:alg:blend}).

\subsection{Segmentation and RR Interval Features}
\label{sec:alg:rpeak}

The digitized ECG samples are segmented into a sequence of heartbeats. The segmentation is performed based on detecting the $R$ peaks\footnote{R peak is a specific point in the ECG waveform as shown in Fig.~\ref{fig:heart}(b).}. 
In specific, every segment (heartbeat) has a fixed length and contains $0.25$~seconds of the input ECG signal before the detected $R$ peak and $0.45$~seconds after. 
This is denoted as $X_{ecg}$ in Fig.~\ref{fig:alg}.

$R$ peak detection algorithms are well established and highly accurate. 
In our segmentation process, Pan-Tompkin's algorithm \cite{Tompkins1985} is used. As part of Pan-Tompkin's algorithm, the time intervals between consecutive $R$ peaks are calculated as well. 
Let $RR_i$ denote the time interval from $R$ peak $i-1$ to $R$ peak $i$. 
Based on this information, we also extract the following four features for heartbeat $i$: 
I) $RR_i$ as the past $RR$ interval, II) $RR_{i+1}$ as the next $RR$ interval, III) $\frac{1}{10}\sum_{k=i-4}^{i+5} RR_k$ as the local average of the five past and the five next $RR$ intervals, and IV) the average duration of the $RR$ intervals in each person's train data. 
These four features are referred to as $RR$ interval features, and form a feature vector denoted as $X_{rr}$ in Fig.~\ref{fig:alg}.

Note that features II and III require access to future heartbeats. However, as opposed to processing previously stored ECG signals, future information is not available in our setting. This is because the proposed algorithm is designed for continuous monitoring. 
We mitigate this problem by buffering the ECG signal in a first-in-first-out (FIFO) memory in real-time. This buffer is implemented in software and must be small. By always classifying the heartbeat which falls in the middle of this buffer, access to the near past and the near future information is made possible.

The fourth feature is the average $RR$ in each person's train data. Section \ref{sec:train:data} discusses the train data. This feature varies among people with different average heart rates. For example, it is larger in athletes because they have slower heart rates.

Besides the above $RR$ interval features which are accurately extracted with minimal computations, the proposed algorithm does not employ other hand-crafted morphological features such as those that are based on $Q$, $S$ or $T$. 
This is because such features are not optimal in representing the characteristics of the underlying signal. 
In addition, they are fixed for all patients under all circumstances and therefore do not efficiently represent the differences among the arrhythmia classes \cite{Hoekema_2001, Kiranyaz2016}. 
Instead, we let the features be automatically extracted using wavelet and recurrent neural networks as discussed in the following sections.

\begin{figure*}[tp]
	\centering
	\includegraphics[width=1.0\textwidth]{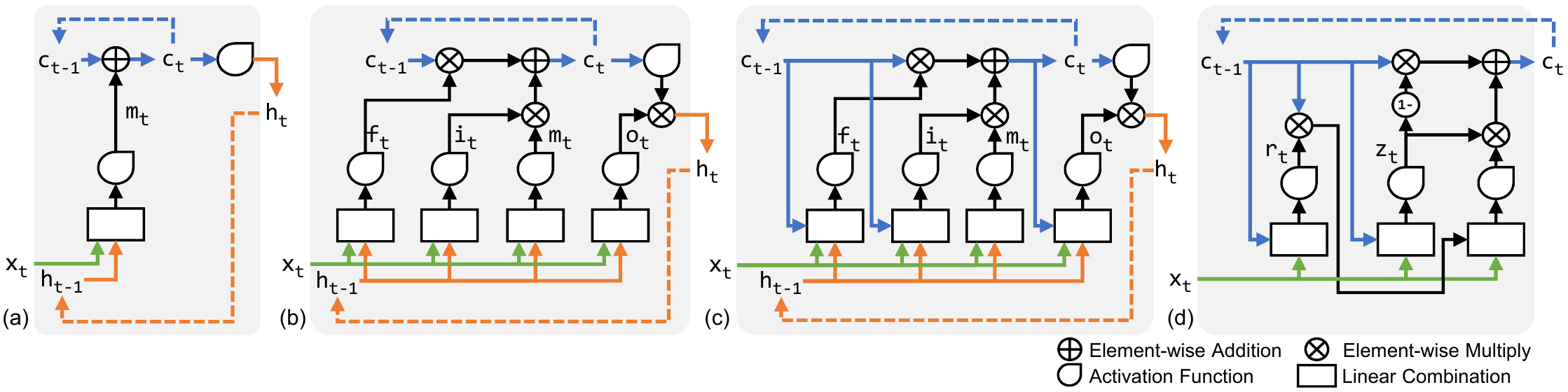} 
	\vskip -1mm
	\caption{(a) Simple RNN Cell, (b) Long Short-Term Memory (LSTM), (c) LSTM with Peepholes, (d) Gated Recurrent Unit (GRU).}
	\label{fig:cells}
	\vskip -2mm
\end{figure*}

\subsection{Wavelet Features}
\label{sec:alg:wavelet}

ECG signal has non-stationary characteristics. Therefore, to capture both the time and the frequency domain information, discrete wavelet transform \cite{Mallat1989} is applied to the digitized ECG samples in every heartbeat. 
In specific, Daubechies wavelet family is selected because of its similarity with the ECG signal \cite{Chazal2000}. 
Low-order Daubechies wavelets have high time resolution but low frequency resolution, while high-order ones have high frequency resolution and low time resolution \cite{Chazal2000}.  
Previous works mostly employed types $1$ to $4$. We employ type $T=2$, i.e., db2, which falls somewhat in the middle of this range, and $L=4$ levels of decomposition. Hence, the final list of wavelet coefficients is $X_w=(\mbox{A4}, \mbox{D4}, \mbox{D3}, \mbox{D2}, \mbox{D1})$.

The computational complexity of discrete wavelet transform on an input array of size $N$ with Daubechies type $T$ and $L$ levels of decomposition is 
\begin{equation}
N \times T \times (1+\frac{1}{2}+\ldots+\frac{1}{2^{L-1}}) 
\label{eq:wavelet:time}
\end{equation}

Since the computational requirement is proportional to the input size, the wavelet input, i.e., $X_{ecg}$ in Fig~\ref{fig:alg}, is down sampled by a factor of $2$ before applying the wavelet transform. 
This down sampling also cuts the total length of the wavelet output for every heartbeat, i.e., $|X_w|$, to about half, and thus helps to reduce the computational requirement of the following steps as well.


\subsection{RNN-based Models}
\label{sec:alg:rnn}

For every heartbeat, the input ECG samples ($X_{ecg}$) along with the extracted $RR$ interval features and wavelet features ($X_{rr}$ and $X_w$) are provided to two separate RNN-based models, called model $\alpha$ and model $\beta$. An overall view of the proposed algorithm is shown in Fig.~\ref{fig:alg}. 
The two models make separate arrhythmia predictions which are then blended to form the final prediction for every heartbeat. 

Employing $X_{rr}$ and $X_w$ in addition to $X_{ecg}$ helps the RNN models capture patterns in the ECG signals more efficiently. The additional features provide processed information to the RNN models. Therefore, accurate results can be reached with smaller and hence faster RNNs. 
Employing multiple smaller RNNs in parallel instead of one larger RNN helps to increase the accuracy without significantly increasing the computational costs. The details are discussed in equation (\ref{eq:time}).

In model $\alpha$, first $X_w$ and $X_{ecg}$ are processed separately and then the outputs are combined.  In model $\beta$, however, $X_w$ and $X_{ecg}$ are first combined and then processed. 
The details are discussed below. 

\vskip 1mm
\subsubsection*{Model $\alpha$}

As shown in Fig.~\ref{fig:alg}, model $\alpha$ consists of two branches. Every branch includes one or two RNNs. Every RNN includes a number of hidden units. 
The input to the left branch is denoted by $X^{\alpha1}$ and is formed by concatenating $X_{rr}$ and $X_{ecg}$. The RNN cells in this branch process the array $X^{\alpha1}$ and extract  $N^{\alpha1}_h$ features. 
Similarly, the right branch concatenates $X_{rr}$ and $X_w$ into array $X^{\alpha2}$, then, processes this array and extracts $N^{\alpha2}_h$ features. 

The outputs of the two branches are concatenated and fed into a fully connected neural network layer in order to produce the probability of all the $N_y$ output arrhythmia classes. The output classes are discussed in Section~\ref{sec:exp}. The dimension of this fully connected layer is equal to $(N^{\alpha1}_h + N^{\alpha2}_h) \times N_y$. The maximum probability determines the arrhythmia class that is predicted by model $\alpha$.

\vskip 1mm
\subsubsection*{Model $\beta$}

As shown in Fig.~\ref{fig:alg}, this model consists of only one branch. As opposed to processing $X_{ecg}$ and $X_w$ in two separate RNN branches and then combining the outputs, here the inputs are combined. 
In specific, array $X^{\beta}$ is formed by concatenating a down sampled version of $X_{ecg}$ with $X_{rr}$ and $X_w$, followed by applying PCA on the concatenated array. 
$X^{\beta}$ is then processed by the RNNs in this model and  $N^{\beta}_h$ features are extracted. The features are fed into a fully connected neural network layer with dimension $N^{\beta}_h \times N_y$. 

Models $\alpha$ and $\beta$ include a number of hyper-parameters, namely, the number of RNNs in every branch, the number of hidden units in every RNN, and the RNN cell types. Hyper-parameter selection is discussed in Section \ref{sec:hyper}. Different RNN cell types are discussed below.

\subsection{RNN Cell Types}
\label{sec:alg:cells}

\subsubsection*{Simple RNN Cell}

Fig.~\ref{fig:cells}(a) shows a simple RNN cell. $x_t$ is the input vector at time $t$. $h_t$ and $c_t$ are state vectors which are carried from time $t-1$ to time $t$, and hence, act as memory by encoding previous information. $h_t$ is also considered as the cell output. 
Size of vectors $h$ and $c$ is denoted by $N_h$ and is known as the number of hidden units. 
The cell works based on the following equations. 
\begin{align}
\label{eq:rnn:m}
&m_{t}[j] =   \\
&\tanh \big( \sum_{k \in [1,N_x]} w[j,k]  x_{t}[k] 
+ \sum_{k \in [1,N_h]} u[j,k]  h_{t-1}[k] 
~+ b[j] \big) \nonumber \\
\label{eq:rnn:c}
&c_{t}[j] = c_{t-1}[j]  + m_{t}[j] \\
\label{eq:rnn:h}
&h_{t}[j] = \tanh(c_{t}[j])
\end{align}

As shown in (\ref{eq:rnn:m}), an intermediate vector $m_t$ is formed by applying $tanh$ activation function on a linear combination of $x_t$ and $h_{t-1}$, i.e., current input and previous output, respectively. $j \in [1,N_h]$. Weight matrices $w$ and $u$ and bias vector $b$ are determined during the training phase (Section \ref{sec:train}). 
The state vector $c_t$ is formed by accumulating $m_t$ over time, as shown in (\ref{eq:rnn:c}). The output vector $h_t$ is formed by applying $tanh$ activation function on $c_t$. 
It can be seen that the output is related to all previous inputs.

\vskip 1mm
\subsubsection*{Long Short-Term Memory (LSTM)}

In the above simple RNN cell the effect of all previous information is accumulated in the internal state vector.  Gradient-based algorithms may fail when temporal dependencies get too long because gradient values may increase or decrease exponentially \cite{Lstm1997}. 

LSTM solves this issue by allowing to forget according to the actual dependencies which exist in the problem. The dependencies are automatically extracted based on the data. This is achieved through forget, input and output gates \cite{Lstm1997}. 
The LSTM cell is shown in Fig. \ref{fig:cells}(b). The gate signals are formed based on $x_t$ and $h_{t-1}$ as shown below. 
\begin{align}
\label{eq:lstm:f}
&f_{t}[j] =   \\
&\sigma \big( \sum_{k \in [1,N_x]} w_f[j,k]  x_{t}[k] 
+ \sum_{k \in [1,N_h]} u_f[j,k]  h_{t-1}[k] 
~+ b_f[j] \big)  \nonumber \\
\label{eq:lstm:i}
&i_{t}[j] =   \\
&\sigma \big( \sum_{k \in [1,N_x]} w_i[j,k]  x_{t}[k] 
+ \sum_{k \in [1,N_h]} u_i[j,k]  h_{t-1}[k] 
~+ b_i[j] \big)  \nonumber \\   
\label{eq:lstm:o}
&o_{t}[j] =   \\
&\sigma \big( \sum_{k \in [1,N_x]} w_o[j,k]  x_{t}[k] 
+ \sum_{k \in [1,N_h]} u_o[j,k]  h_{t-1}[k] 
~+ b_o[j] \big) \nonumber
\end{align}

In the above equations, $\sigma$ denotes the $sigmoid$ activation function, and $j \in [1,N_h]$. 
In the LSTM cell, $m_t$ is computed as before, i.e., as in (\ref{eq:rnn:m}), but (\ref{eq:rnn:c}) and (\ref{eq:rnn:h}) are modified based on the forget, input and output gate signals as the following. 
\begin{align}
\label{eq:lstm:c}
&c_{t}[j] = f_t[j] \times c_{t-1}[j]  + i_t[j] \times m_{t}[j]  \\
\label{eq:lstm:h}
&h_{t}[j] = o_t[j] \times \tanh(c_{t}[j])
\end{align}

As shown in (\ref{eq:lstm:c}), the forget gate $f_t$ controls carrying of state vector $c$ from time $t-1$ to time $t$. The input gate $i_t$ adjusts the accumulation of $m_t$ in $c_t$. 
As shown in (\ref{eq:lstm:h}), the output $h_t$ is formed by applying $tanh$ activation function on $c_t$, and is then adjusted by the output gate $o_t$.

As the above equations show, the LSTM output still depends on all previous inputs. 
Previous information is neither completely discarded nor completely carried over to the current state. 
Instead, influence of the previous information on the current state is carefully controlled through the gate signals \cite{Lstm1997}.

\vskip 1mm
\subsubsection*{LSTM with Peepholes}

The LSTM cell can be extended by adding extra connections from the internal state vector to the forget, input and output gates. The extra connections are marked with blue color in Fig. \ref{fig:cells}(c). The gate signals are formed based on a linear combination of $x_t$, $h_{t-1}$ and now also $c_{t-1}$ \cite{Peephole}. Detailed equations are omitted for brevity.  

\vskip 1mm
\subsubsection*{Gated Recurrent Unit (GRU)}

This cell is a simplified version of the LSTM cell which merges the two state vectors into one and also employs a different gating strategy \cite{Gru2014}. The GRU cell is shown in Fig. \ref{fig:cells}(d). Here, $c_t$ is formed as 
\begin{align}
\label{eq:gru:c}
&c_{t}[j] = (1-z_t[j]) \times c_{t-1}[j] ~+~ z_t[j] \times   \\
&\tanh( \sum_{k \in [1,N_x]} w[j,k]  x_{t}[k] 
+ \sum_{k \in [1,N_h]} u[j,k]  r_t[k] c_{t-1}[k] 
+ b[j] ) \nonumber 
\end{align}
where, $z_t$ and $r_t$ are update and reset gate signals, respectively, and are formed similar to the LSTM gate signals as linear combinations of $x_{t}$ and $c_{t-1}$. 

\vskip 1mm
\subsubsection*{Complexity Analysis}

The above RNN cells perform several matrix and vector operations. 
For instance, the LSTM cell requires four matrix vector multiplications of size $N_h \times N_x$, four matrix vector multiplications of size $N_h \times N_h$ and several vector operations of size $N_h$. 
Total computational complexity for every execution of an RNN cell is therefore equal to 
\begin{equation}
a N_x N_h + b N^2_h + c N_h + d
\label{eq:time}
\end{equation}
where $a$, $b$, $c$ and $d$ depend on the cell type. 
According to the above equation, the computational complexity of an  RNN cell has a quadratic growth with respect to the number of hidden units, i.e., $N_h$. Therefore, multiple smaller RNNs have lower computational costs in total compared to one larger RNN. 
For instance, the total runtime of two RNNs with $N_h=X$ is smaller than one RNN with $N_h=2X$. 



\subsection{Blend Model}
\label{sec:alg:blend}

Ensemble methods such as blending are designed to boost the classification accuracy by blending the predictions made by multiple learning models \cite{Sill2009}. 
As shown in Fig.~\ref{fig:alg}, only two models are blended in our proposed algorithm in order to keep the computational requirement as low as possible. 
For every heartbeat, first, the two RNN-based models $\alpha$ and $\beta$ independently compute the probability of all the $N_y$ output arrhythmia classes. 
Then the two results are blended to form the final probability of the $N_y$ output classes.

The blend model is implemented using a multi-level perceptron (MLP) with two hidden layers. 
The input and output layers have $2 \times N_y$ and $N_y$ neurons, respectively. 

\begin{figure}[tp]
	\centering
	\includegraphics[width=1.0\columnwidth]{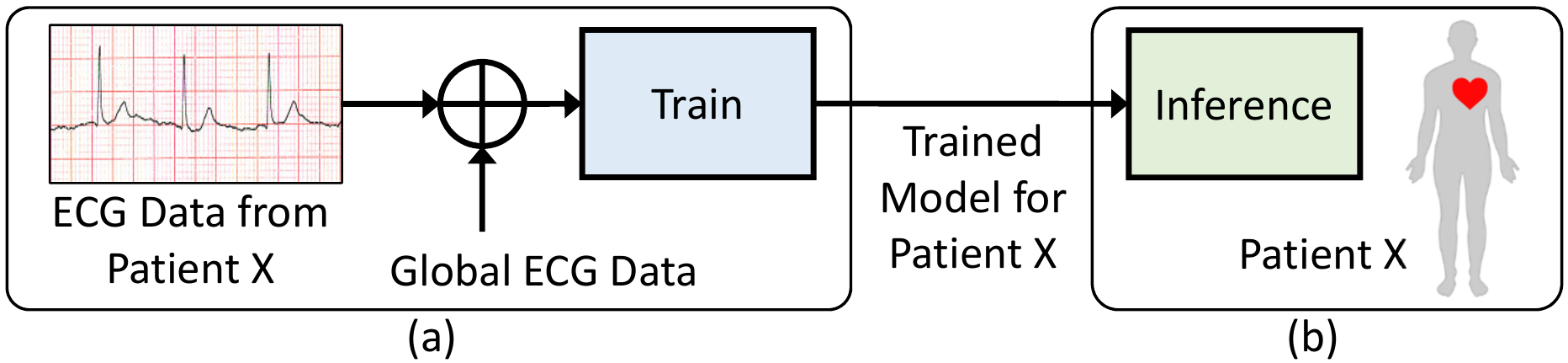}
	\vskip -1mm
	\caption{(a) Patient-specific training. (b) Continuous ECG monitoring and heartbeat classification in real-time.}
	\vskip -2mm
	\label{fig:train}
\end{figure}

\section{Training Procedure}
\label{sec:train}

\subsection{Patient-Specific Training}
\label{sec:train:data}

We employ a patient-specific training procedure. In other words, the model is trained for every patient individually  \cite{Hu1997,Chazal2006,Jiang2007,Ince2009,Kiranyaz2016}. 
Once the model is trained for a patient, continuous ECG monitoring and heartbeat classification is performed in real-time based on the trained model of that patient. This is shown in Fig.~\ref{fig:train}. 
Note that training is performed only once for every patient, i.e., it is not performed continuously.

As shown in Fig.~\ref{fig:train}(a), the training data for a patient is formed by combining two sets of data: local ECG data and global ECG data. 
The first part, i.e., local data, is specific to the patient and is helpful in increasing the classification accuracy due to existing similarities among the heartbeats of every patient. According to AAMI standards \cite{AAMI}, this ECG data can be at most five minutes long. 
The second part, i.e., global data, is the same for all patients. It consists of a number of representative heartbeats from all arrhythmia classes. It helps the model learn other arrhythmia patterns that are not included in the local data. 
Details of the ECG signals employed in our experiments are presented in Section \ref{sec:exp}.

Patient-specific training has been employed in \cite{Hu1997,Chazal2006,Jiang2007,Ince2009,Kiranyaz2016} as well. 
Another approach is to train only one model by feeding data from many patients, and then, use the trained model for classification of data from other patients. 
We do not employ this approach because the ECG waveform varies significantly among different patients \cite{Kiranyaz2016}. 

\subsection{Train the RNN Models}
\label{sec:train:rnn}

Back propagation (BP) is a well known method for training feed-forward neural networks such as convolutional neural networks (CNNs). This method cannot be applied to RNNs because of the existing temporal dependencies in the model, i.e., the feedback loops in Fig.~\ref{fig:cells} which carry previous information through time. 

To train RNN models, train data is split into batches of several heartbeats each. The heartbeats are processed sequentially as the following. 
The weights are updated upon completion of every batch. 
In the beginning of every batch, $h$ is set to zero and $c$ is set randomly. 
Then the input data is forward-propagated over the network, and error is calculated until the batch finishes. Next, the error is back-propagated over the unfolded network in time, the weight matrices change in all instances and their mean is set as the updated weight. This is repeated until all batches are processed. This method is known as back propagation through time (BPTT) \cite{Werbos1988}. 
The optimization method employed in our work is adaptive moment estimation algorithm (Adam).

\subsection{Train the Blend Model}
\label{sec:train:blend}

First, the two RNN-based models, i.e., model $\alpha$ and model $\beta$, are trained independently as discussed above. Next, their output arrhythmia predictions for the heartbeats in the train data are used to train the blend model, i.e., the multi-level perceptron in Fig.~\ref{fig:alg}. The training is performed using back propagation (BP). 
 

\subsection{Hyper-Parameter Selection} 
\label{sec:hyper}

Learning algorithms related to neural networks often involve hyper-parameters. There are a number of guidelines and recommendations for selecting the hyper-parameters \cite{Bengio2012}. 
For the RNN-based models $\alpha$ and $\beta$, we perform a grid search on the range of $1 - 2$ for the number of recurrent layers, i.e., the number of RNNs in every branch, and $10 - 200$ for the number of hidden units, i.e., $N^{\alpha1}_h$, $N^{\alpha2}_h$ and $N^{\beta}_h$. 
For the cell type, we consider simple RNN cell, LSTM, LSTM with peephole, and GRU. 
Note that hyper-parameter selection is performed independently for model $\alpha$, model $\beta$ and the blend model. 
We found that RNN models with the LSTM cell, $N^{\alpha1}_h=30$, $N^{\alpha2}_h=30$, $N^{\beta}_h=50$ and one recurrent layer achieve consistently strong results.

\section{Experimental Results}
\label{sec:exp}

\subsection{Setup and ECG Data}
\label{sec:exp:data}

The proposed algorithm is implemented in the Python language and TensorFlow \cite{TensorFlow} library. 
%
%
Our source code is available online \cite{SourceCode}.

MIT-BIH ECG arrhythmia database \cite{MIT-BIH} is used to evaluate the proposed algorithm and compare its performance with previous works. 
Each record in this database has two leads. 
The first lead is modified limb lead II. The second one is modified lead V1 or in some cases V2, V4 or V5.
Two or more cardiologists independently annotated each record. 
The database contains two sets of data, called DS100 and DS200. 
DS100 includes representative samples of the variety of ECG waveforms and artifacts that an arrhythmia detector might encounter in routine clinical practice. 
DS200 includes complex ventricular, junctional, and supraventricular arrhythmias and conduction abnormalities. 
Based on AAMI standards, the records that contain paced beats ($102,104,107,217$) are excluded  \cite{AAMI}.

Training procedure is discussed in Section \ref{sec:train:data}. The model is individually trained for every patient. Two sets of data, namely local data and global data, are combined for training the model for every patient. 
Global train data is formed by randomly selecting representative heartbeats from all arrhythmia classes in DS100 records.  
Local train data is the first five minutes of a patient's record in DS200. This is in compliance with AAMI standards \cite{AAMI}. 
Test data is all the records in DS200. The first five minutes of all the records are skipped in the test data. 

\begin{table}[tp]
	\centering
	\includegraphics[width=1.0\columnwidth]{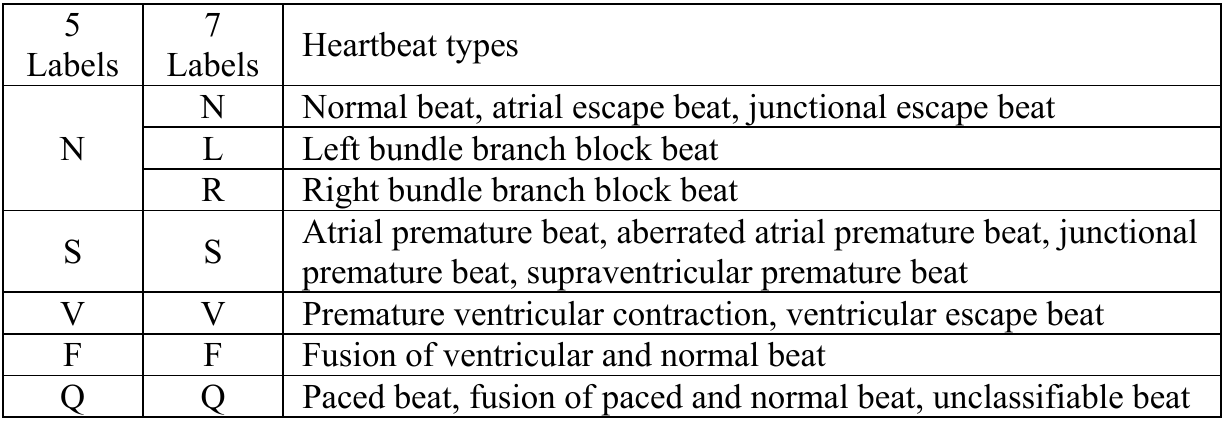}
	\vskip -1mm
	\caption{Heartbeat classes.}
	\label{fig:labels}
	\vskip 1mm
	\includegraphics[width=1.0\columnwidth]{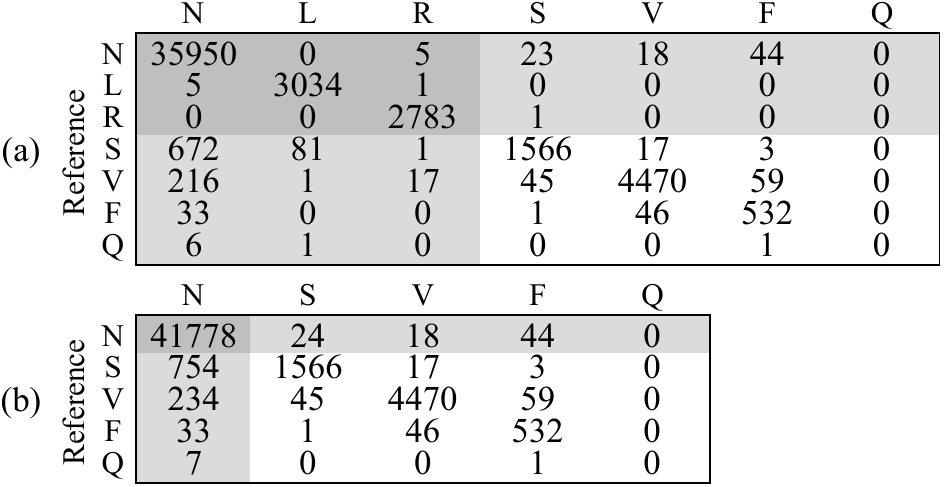}
	\vskip -1mm
	\caption{Confusion matrix with (a) $7$ and (b) $5$ heartbeat classes.}
	\vskip -2mm
	\label{fig:CM57}	
\end{table}


\begin{table*}[tp]
	\centering
	\includegraphics[width=1.0\textwidth]{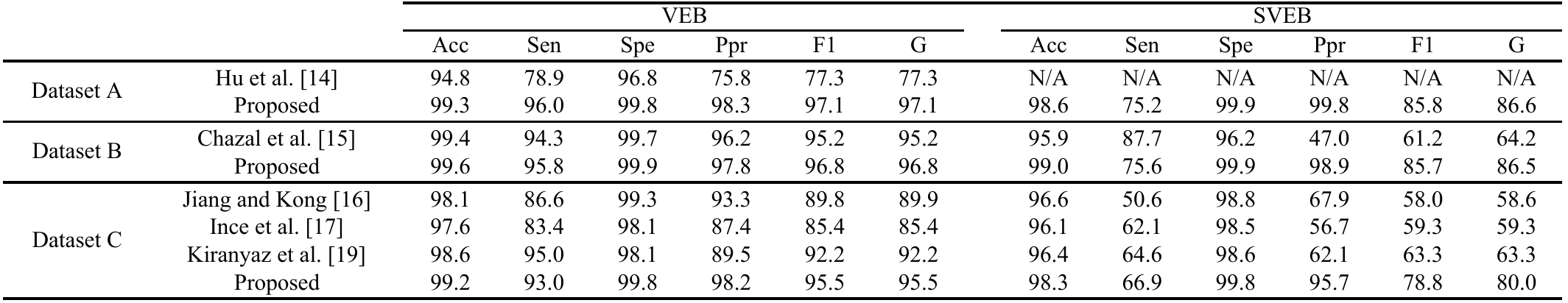}
	\vskip -1mm
	\caption{
		Comparing the proposed algorithm with previous works in binary classification of VEB and binary classification of SVEB. 
		Dataset A for VEB classification is 200, 202, 210, 213, 214, 219, 221, 228, 231, 233 and 234. Dataset A for SVEB classification is the same records for VEB classification plus 212, 222 and 232 \cite{Hu1997}. 
		Dataset B is 100, 103, 105, 111, 113, 117, 121, 123, 200, 202, 210, 212, 213, 214, 219, 221, 222, 228, 231, 232, 233 and 234 \cite{Chazal2006}. 
		Dataset C is 200, 201, 202, 203, 205, 207, 208, 209, 210, 212, 213, 214, 215, 219, 220, 221, 222, 223, 228, 230, 231, 232, 233 and 234 \cite{Jiang2007,Ince2009,Kiranyaz2016}.
	}
	\vskip -2mm
	\label{fig:Compare}
\end{table*}

\subsection{Classification Performance}
\label{sec:exp:accuracy}

In our experimental evaluations, every heartbeat is classified into the seven arrhythmia classes that are shown in Table~\ref{fig:labels}. Based on AAMI standards \cite{AAMI}, many previous works employ five class labels, namely, N, S, V, F and Q \cite{Jiang2007,Ince2009,Kiranyaz2016}. However, to have more resolution, we split class N into three classes by separating two conduction abnormalities known as left bundle branch block (L) and right bundle branch block (R). 
As shown in Table~\ref{fig:CM57}(a), the proposed algorithm is able to distinguish L and R from N very efficiently. 
To compare the proposed algorithm with previous works, L and R are merged back into N as shown in Table~\ref{fig:CM57}(b).

In order to report performance results for binary classification of ventricular ectopic beats (VEB) from non-VEBs and also supraventricular ectopic beats (SVEB) from non-SVEBs, four statistical metrics, namely, accuracy ($Acc$), sensitivity ($Sen$), specificity ($Spc$), and positive predictivity ($Ppr$) are extracted from the confusion matrix.  
\begin{align}
&Acc = \frac{\scriptstyle TP+TN}{\scriptstyle TP+TN+FP+FN} \displaystyle \\
&Sen = \frac{\scriptstyle TP}{\scriptstyle TP+FN} \displaystyle \\
&Spe = \frac{\scriptstyle TN}{\scriptstyle TN+FP} \displaystyle \\
&Ppr = \frac{\scriptstyle TP}{\scriptstyle TP+FP} \displaystyle
\end{align}

The terms TP, TN, FP and FN denote true positive, true negative, false positive and false negative in the binary classification, respectively. 
Since, increasing $Ppr$ often decreases $Sen$ and vice versa, $F_1$ and $G$ scores are also calculated which combine $Sen$ and $Ppr$ as the following. 
\begin{align}
&F_1 = \frac{\scriptstyle 2}{\frac{\scriptstyle 1}{\scriptstyle Sen}+\frac{\scriptstyle 1}{\scriptstyle Ppr}} \\
&G = \sqrt{ \scriptstyle Sen \times Ppr}
\end{align}

Table~\ref{fig:Compare} compares the proposed algorithm with previous works. 
In order to provide thorough and fair comparisons, we employ the exact same data as the previous works. 
The proposed algorithm cannot be directly compared with other methods that do not comply with AAMI standards or do not employ the openly available MIT-BIH database. 
To compare the proposed algorithm with the methods in \cite{Hu1997} and \cite{Chazal2006} we consider datasets A and B, respectively. The main dataset that is also employed in \cite{Jiang2007}, \cite{Ince2009} and \cite{Kiranyaz2016} is dataset C.

In both VEB and SVEB, the proposed algorithm achieves superior classification performance compared to previous works. 
In VEB detection, for instance, accuracy is always higher than $99\%$. It is $4.5\%$, $0.2\%$ and $0.6\%$ higher than the previous works in datasets A, B and C, respectively. 
$F_1$ score is much higher. It is $19.8\%$, $1.6\%$ and $3.3\%$ higher than the previous works in datasets A, B and C, respectively. 

In SVEB detection, accuracy is $3.1\%$ and $1.7\%$ higher than the previous works in datasets B and C, respectively. 
$F_1$ score is $24.5\%$ and $15.5\%$ higher than the previous works in datasets B and C, respectively. 
Note that $F_1$ score is a more meaningful metric compared to accuracy.


\subsection{Real-time Execution}
\label{sec:exp:time}  

Personal wearable devices have small and low-power processors which are much slower compared to desktop and server processors. Therefore, to meet timing requirements for continuous execution, the proposed heartbeat classification algorithm needs to have low computational intensity. 

\begin{figure}[tp]
	\centering
	\includegraphics[width=1.0\columnwidth]{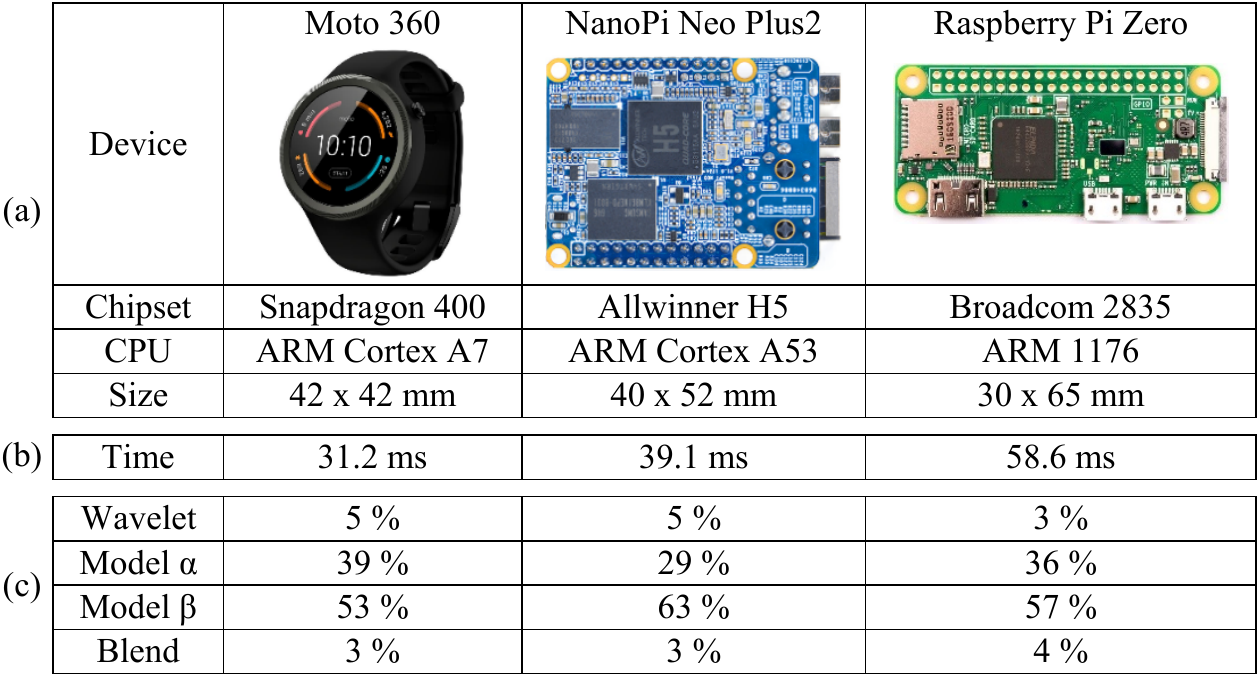}
	\vskip -1mm	
	\caption{a) Hardware platforms. b) Measured execution time. c) Distribution of the execution time.}
	\vskip -2mm	
	\label{fig:platform}
\end{figure}

Note that it is only the inference (test) phase which is executed repeatedly in real-time and needs to meet timing requirements. The training phase is performed only once in the beginning. 
In this section, we experimentally evaluate the execution time of the test phase on the hardware platforms shown in Fig.~\ref{fig:platform}(a). 
All these platforms have small and low-power processors. 
The Java language and Android Studio are used to implement the source code for Moto 360 which is an AndroidWear device, and the C language is used for the other two hardware platforms.

Measured execution times are shown in Fig.~\ref{fig:platform}(b). The proposed algorithm takes about $30$ to $60$ milliseconds to classify every heartbeat, while assuming a maximum heart rate of $200$ bpm, a time window of at least $300$ milliseconds is available. 
This shows that the proposed algorithm meets timing requirements for continuous ECG classification on small and low-power hardware platforms. 

\section{Discussion}
\label{sec:exp:dis}

\subsection{Ablation Study}

In this section, different parts of the proposed model are modified and the results are experimentally studied. This helps to provide a better understanding of the impact of different parts of the proposed model.

\vskip 1mm
\subsubsection*{No Wavelet} 

In order to experimentally study the effect of the wavelet features, we perform the same experiments described above but without the wavelet. 
Fig.~\ref{fig:cmp} compares the results against the results of the original solution, i.e., the proposed solution. In specific, it shows the amount of degradation in the $F_1$ score. 

We see that removing the wavelet features reduces the $F_1$ score by $5.1\%$ and $8.3\%$ for VEB and SVEB detection, respectively. 
This is because the wavelet transform provides processed information to the LSTM models, and thus, helps the models learn different patterns more efficiently. 
In addition, note that as shown in Fig.~\ref{fig:platform}(c), the wavelet transform adds a very small overhead to the overall execution time. 

\vskip 1mm
\subsubsection*{Wavelet Types} 

We experiment with different wavelet types as well. Here db2 (the selected type) is replaced with db1, db3 and db4. As shown in Fig.~\ref{fig:cmp}, db2 and db3 which fall in the middle of this range yield the best performance, but db1 and db4 degrade the $F_1$ score. 
This is expected because higher time resolution is achieved by low-order types, while higher frequency resolution is achieved by high-order types. 
Between db2 and db3, we selected db2 because it employs a very small $4$-point convolution kernel, and thus, provides a more computationally lightweight configuration.

\vskip 1mm
\subsubsection*{RNN Cell Types}

Next, the effect of using different RNN cell types is studied. In specific, the employed LSTM cell is replaced with simple RNN, GRU and Peephole cells. The results are shown in Fig.~\ref{fig:cmp}. 
The simple RNN and GRU cells degrade the classification performance, but the Peephole cell is very close to LSTM. We employ LSTM because it does not have the extra computations required in the Peephole cell, i.e., the extra blue connections in Fig.~\ref{fig:cells}(c).

\vskip 1mm
\subsubsection*{No Blending (Model $\alpha$)} 

\begin{figure}[tp]
	\centering
	\includegraphics[width=1.0\columnwidth]{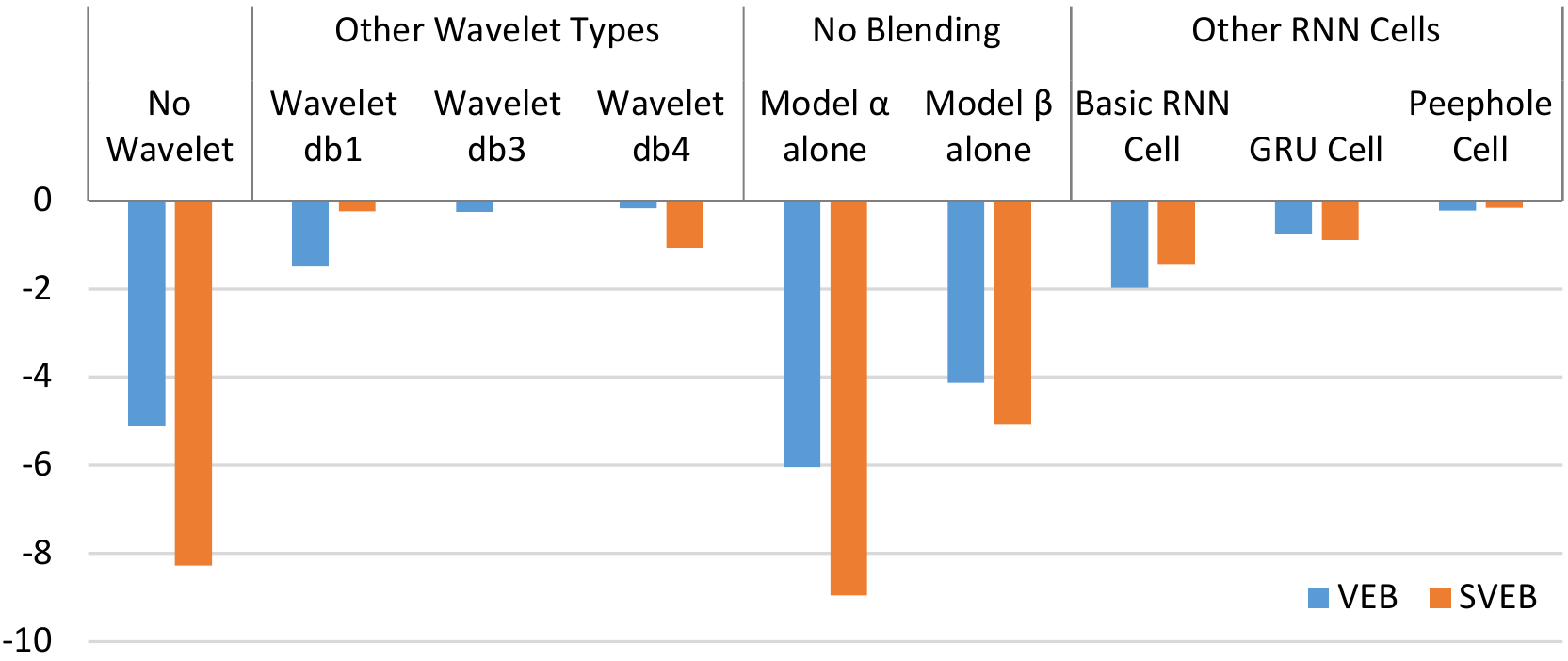}
	\vskip -1mm	
	\caption{Degradation of the $F_1$ score when different parts of the proposed algorithm are modified.}
	\vskip -2mm	
	\label{fig:cmp}
\end{figure}

In order to study the effect of combining two models, we perform the same experiments but with only one of the two models, in specific, model $\alpha$. Hence, in addition to removing model $\beta$, the blend model itself is also removed. 
Fig.~\ref{fig:cmp} compares the results of this experiment with the proposed solution in which models $\alpha$ and $\beta$ are blended. 
When only model $\alpha$ is present, $F1$ score is $6\%$ and $9\%$ lower in VEB and SVEB detection, respectively. This shows that blending highly increases the classification performance.

\vskip 1mm
\subsubsection*{No Blending (Model $\beta$)} 

Similarly, when only model $\beta$ is present, the classification performance is degraded. In specific, $F1$ score is $4.1\%$ and $5.1\%$ lower in VEB and SVEB detection, respectively. 

Fig.~\ref{fig:platform}(c) shows the distribution of the execution time in our hardware platforms.  Model $\beta$ has longer execution time compared to model $\alpha$. This is because it has a larger LSTM cell. In specific, $N^{\beta}_h=50$ but $N^{\alpha1}_h=N^{\alpha2}_h=30$. As discussed in details in equation (\ref{eq:time}), two smaller LSTMs have lower computational costs in total compared to one larger LSTM.

\subsection{Classification Performance with Limited Data} 

\subsubsection*{Single ECG Lead} 

The above experiments are based on two ECG leads. This is similar to previous works. However, in some wearable health monitoring devices, only one lead is available. Here the proposed algorithm is re-evaluated based on data from the first lead. 
The results are shown in Fig.~\ref{fig:limited_data}. The proposed algorithm shows a relatively lower but still acceptable classification performance in this setting.

\vskip 1mm
\subsubsection*{2.5 Minutes ECG Data} 

\begin{figure}[tp]
	\centering
	\includegraphics[width=1.0\columnwidth]{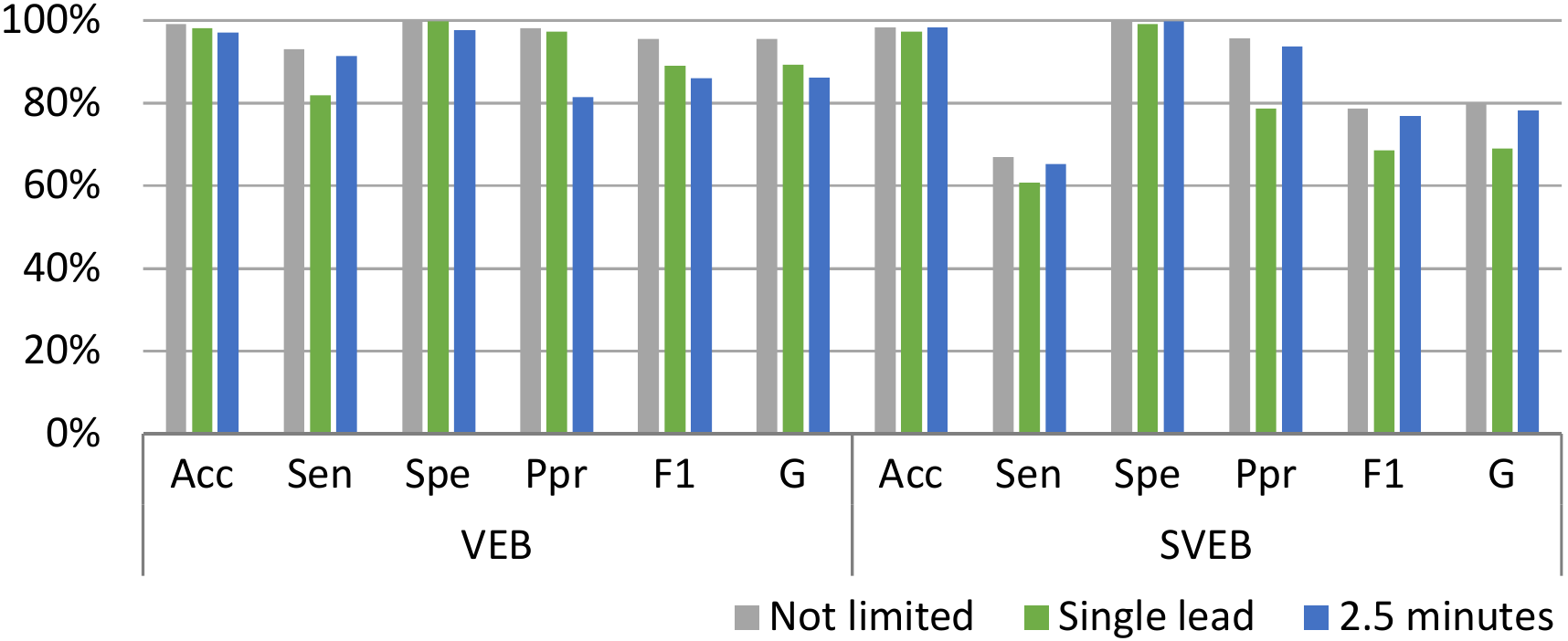}
	\vskip -1mm	
	\caption{Classification performance with limited ECG data.}
	\vskip -2mm	
	\label{fig:limited_data}
\end{figure}

Similar to previous patient-specific methods, the first $5$ minutes of a patient's ECG record is used here as local data in the training phase. 
In practical settings, this data needs to be visually inspected and labeled by a specialist. Cutting the length of this data to half can help reduce the associated time and costs. 
Fig.~\ref{fig:limited_data} shows the classification performance of the proposed algorithm when local data is only $2.5$ minutes long. 
The classification performance is relatively lower but still acceptable.

\section{Conclusion}
\label{sec:conc}

In this paper a novel LSTM-based ECG classification algorithm was proposed which achieves superior classification performance compared to previous works. In addition, as opposed to many previous deep-learning based algorithms, it has low computational costs and meets timing requirements for continuous execution on wearable devices with limited processing power. 
Future directions include exploring other techniques to further increase the classification performance, studying other features in addition to wavelet, and improvements on single-lead ECG processing.

\ifCLASSOPTIONcaptionsoff
  \newpage
\fi



%
\bibliographystyle{IEEEtran}


%

%
%
%




\end{document}